\documentclass[reprint,superscriptaddress,
prb,aps,
]{revtex4-2}
\setcitestyle{super} 
\usepackage{graphicx}
\usepackage{dcolumn}
\usepackage{bm}
\usepackage[version=4]{mhchem}
\usepackage{ragged2e}
\usepackage{booktabs}

\usepackage{siunitx}
\DeclareSIUnit\bar{bar}
\usepackage{hyperref}
\hypersetup{
            hypertex=true,
            colorlinks=true,
            linkcolor=blue,
            anchorcolor=blue,
            citecolor=blue}

\usepackage{pdfpages}

\makeatletter
\AtBeginDocument{\let\LS@rot\@undefined}
\makeatother

\usepackage{pgffor}

\begin{document}

\title{Finite Element-based Nonlinear Dynamic Optimization of Nanomechanical Resonators}

\author{Zichao Li}
\email[]{z.Li-16@tudelft.nl}
\affiliation{Faculty of Mechanical Engineering, Department of Precision and Microsystems Engineering, Delft University of Technology, Mekelweg 2, 2628 CD Delft, The Netherlands}
\author{Farbod Alijani}
\author{Ali Sarafraz}
\affiliation{Faculty of Mechanical Engineering, Department of Precision and Microsystems Engineering, Delft University of Technology, Mekelweg 2, 2628 CD Delft, The Netherlands}
\author{Minxing Xu}
\affiliation{Faculty of Mechanical Engineering, Department of Precision and Microsystems Engineering, Delft University of Technology, Mekelweg 2, 2628 CD Delft, The Netherlands}
\affiliation{Kavli Institute of Nanoscience, Delft University of Technology, Lorentzweg 1, 2628 CJ Delft, The Netherlands}
\author{Richard A. Norte}
\affiliation{Faculty of Mechanical Engineering, Department of Precision and Microsystems Engineering, Delft University of Technology, Mekelweg 2, 2628 CD Delft, The Netherlands}
\affiliation{Kavli Institute of Nanoscience, Delft University of Technology, Lorentzweg 1, 2628 CJ Delft, The Netherlands}
\author{\\Alejandro M.~Aragón}
\affiliation{Faculty of Mechanical Engineering, Department of Precision and Microsystems Engineering, Delft University of Technology, Mekelweg 2, 2628 CD Delft, The Netherlands}
\author{Peter G. Steeneken}
\email[]{p.g.steeneken@tudelft.nl}
\affiliation{Faculty of Mechanical Engineering, Department of Precision and Microsystems Engineering, Delft University of Technology, Mekelweg 2, 2628 CD Delft, The Netherlands}
\affiliation{Kavli Institute of Nanoscience, Delft University of Technology, Lorentzweg 1, 2628 CJ Delft, The Netherlands}


\begin{abstract}
Nonlinear dynamic simulations of mechanical resonators have been facilitated by the advent of computational techniques that generate nonlinear reduced order models (ROMs) using the finite element (FE) method. However, designing devices with specific nonlinear characteristics remains inefficient since it requires manual adjustment of the design parameters and can result in suboptimal designs. Here, we integrate an FE-based nonlinear ROM technique with a derivative-free optimization algorithm to enable the design of nonlinear  mechanical resonators. The resulting methodology is used to optimize the support design of high-stress nanomechanical \ce{Si3N4} string resonators, in the presence of conflicting objectives such as simultaneous enhancement of $Q$-factor and nonlinear Duffing constant. To that end, we generate Pareto frontiers that highlight the trade-offs between optimization objectives and validate the results both numerically and experimentally. To further demonstrate the capability of multi-objective optimization for practical design challenges, we simultaneously optimize the design of nanoresonators for three key figure-of-merits in resonant sensing: power consumption, sensitivity and response time. The presented methodology can facilitate and accelerate designing (nano)mechanical resonators with optimized performance for a wide variety of applications.



\end{abstract}

\maketitle


\newpage

\section{Introduction}

Design of mechanical structures that move or vibrate in a predictable and desirable manner is a central challenge in many engineering disciplines. This task becomes more complicated when these structures experience large-amplitude vibrations, since linear analysis methods fail and nonlinear effects need to be accounted for. This is particularly important at the nanoscale, where forces on the order of only a few pN can already yield a wealth of nonlinear dynamic phenomena worth exploiting~\cite{HanayPRL2020,HanayNanoLett2020,KoleySwitch2022, venstra2013stochastic,miller2021amplitude}.  



Although design optimization of micro and nanomechanical resonators in the linear regime is well-established~\cite{hoj2021ultra}, the use of design optimization for engineering nonlinear resonances has received less attention~\cite{schiwietz2024shape}. This is because designers tend to avoid the nonlinear regime, and optimizing structures' nonlinear dynamics is more complex, which requires extensive computational resources. As a result, available literature on nonlinear dynamic optimization is limited, although some recent advances have been made that combine analytical methods with gradient-based shape optimization, to optimize nonlinearities in micro beams~\cite{Dou2015, LilyLi}. For nonlinear modeling of more complex structures, several approaches have been developed based on nonlinear reduced order modeling (ROM) of finite element (FE) simulations~\cite{mignolet2013review, touze2021model, cenedese2022data}. A particularly attractive class known as STEP (STiffness Evaluation Procedure) ~\cite{muravyov2003determination} can determine nonlinear coefficients of an arbitrary mechanical structure and can be implemented in virtually any commercial finite element method (FEM) package. This, for instance, has been recently shown by using COMSOL to model the nonlinear dynamics of high-stress \ce{Si3N4} string~\cite{li2024strain} as well as graphene nanoresonators~\cite{AtaPhysRevApplied}. Since the number of degrees of freedom in the ROM is much smaller than that in the full FE model, the nonlinear dynamics of the structure can be simulated much more rapidly using numerical continuation packages~\cite{Matcont}. 

In this work, we present a route for nonlinear dynamic optimization that is based on an FE-based ROM. The methodology, which is a combination of Particle Swarm Optimization (PSO) with STEP~\cite{muravyov2003determination} (OPTSTEP), has several beneficial features. First of all, because it uses a derivative-free optimization routine for approaching the optimal design, it can be implemented and combined with FEM packages that are not able to obtain gradients easily. Secondly, the ROM parameters generated in OPTSTEP can facilitate explicitly expressing the optimization goals. Finally, as will be shown, the developed procedure allows using multiple objective functions to approximate a Pareto front, which can help designers in decision-making processes when having to balance performance trade-offs among different objectives. Considering the outstanding performance as ultrasensitive mechanical detectors and the mature fabrication procedure~\cite{xu2024high,cupertino2024centimeter}, we select high-stress \ce{Si3N4} for the experimental validation of our methodology.

The manuscript is structured as follows. We first introduce and describe the general OPTSTEP methodology. Then we demonstrate the method on the specific challenge of the optimization of the support structure for a high-stress \ce{Si3N4} nano string, while taking the maximization of its $Q$-factor and nonlinear Duffing constant $\beta$ as examples of linear and nonlinear objectives. By comparing the PSO results to the $Q$ and $\beta$ values that result from a brute-force simulation of a large number of designs that span the design space, we validate that OPTSTEP finds the optimum designs much faster with the same computational resources. 
Subsequently, we turn to the problem of dealing with multiple objective functions and focus on simultaneously maximizing both $Q$ and $\beta$, demonstrated by a Pareto front. For validation, the results are compared to experimental measurements of fabricated devices. We conclude by demonstrating the potential of OPTSTEP for optimizing the performance of resonant sensors by using more complex objective functions that are relevant for engineering their response time, sensitivity, and power consumption. \\[1pt]

\begin{figure*}
\includegraphics[scale=0.88
]{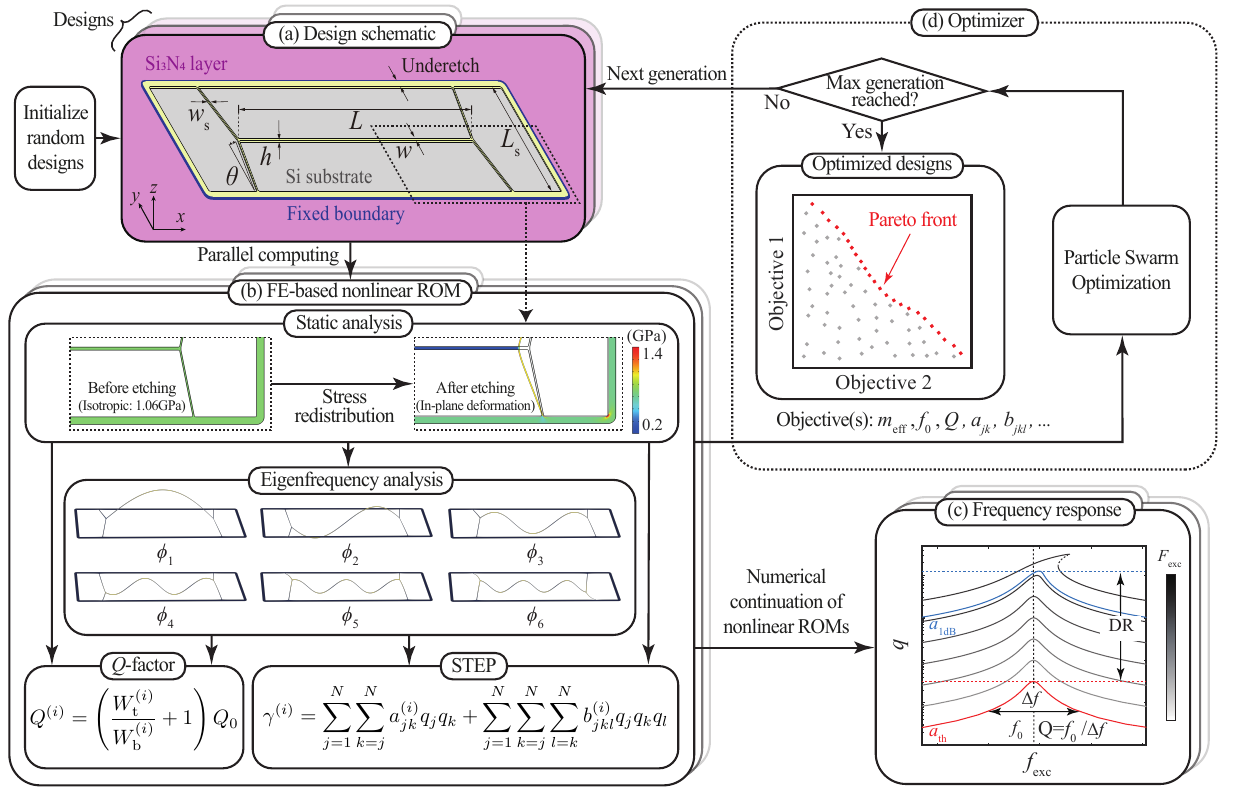}
\caption{\label{fig.1} \textbf{Schematic of the OPTSTEP method} (a) A device geometry is chosen and parameterized by a set of design optimization variables.  In this specific case a \ce{Si3N4} nanomechanical string resonator is chosen for demonstrating OPTSTEP. (b) All designs in one generation are simulated in parallel on a high-performance computing cluster. Static analysis is conducted to evaluate the stress redistribution and deformation after etching, followed by eigenfrequency analysis. Resonance frequencies, mode shapes, Q-factor and the ROM are obtained from the full FE model. (c) The ROM is simulated by numerical continuation. (d) Objective(s) selected from ROM are sent to an optimizer (PSO in this study) to generate design variables for the next generation. 
}
\end{figure*}

\section{OPTSTEP methodology}
An overview of the OPTSTEP method is schematically shown in Fig.~\ref{fig.1}. In the current work, we use it for engineering a parameterized geometry. We use nanomechanical string resonators with compliant supports, which is shown in Fig.~\ref{fig.1}a, to demonstrate the methodology. We keep the length $L$ and width $w$ of the central string constant, while varying the width $w_{\rm s}$, length $L_{\rm s}$ and angle $\theta$ of the supports, as well as the thickness $h$ of the device. It is noted that the OPTSTEP methodology might be used with a larger number of parameters, or even might be extended towards shape or topology optimization of nonlinear dynamic structures. However, such extension is out of the scope of the current work.

For a certain set of geometrical parameters, a ROM for the parameterized structure is generated using the STEP method~\cite{muravyov2003determination}, which we implemented with shell elements in COMSOL~\cite{li2024strain}. Besides geometric parameters and boundary conditions (see Fig.~\ref{fig.1}a), the COMSOL simulation contains material parameters (see Methods), and the initial pre-stress distribution is calculated using a static analysis~\cite{li2024strain}. We conduct this static analysis assuming the material is isotropic and  pre-stressed ($\sigma_0=\qty[]{1.06}{\giga\pascal}$). We then calculate the stress redistribution during the sacrificial layer underetching process, whereby the high-stress \ce{Si3N4} layer releases from the silicon substrate. Note that in the present study we only consider $\theta \geq 0$, such that the central string is always in tension (in contrast to Ref.~\citenum{li2024strain}). After the static analysis, an eigenfrequency analysis is performed to obtain the out-of-plane eigenmodes $\phi_i$ (see Fig.~\ref{fig.1}b). These eigenmodes, together with the redistributed stress field obtained from the static analysis, are then used to determine the $Q$-factor, resonance frequency $f_0$, and effective mass $m_\mathrm{eff}$, following the procedure outlined in Ref.~\citenum{li2023tuning}.

As indicated in Fig.~\ref{fig.1}b the STEP method generates a set of coupled nonlinear differential equations\cite{muravyov2003determination, li2024strain, AtaPhysRevApplied}, where the effective nonlinear elastic force acting on the $i$th mode is given by the function $\gamma^{(i)}$ that depends on the quadratic $a_{ij}$, cubic $b_{ijk}$ coupling coefficients, and the generalized coordinates $q_i$. $q_i$ describes the instantaneous contribution of the corresponding mode shapes $\phi_i$ to the deflection of the structure.  
Thus, the finite element model with several thousand or even millions of degrees of freedom (DOFs) is reduced to a condensed ROM, that can usually describe the nonlinear dynamics to a good approximation with less than ten degrees of freedom. We can visualize the resulting frequency response curves for different harmonic drive levels by numerical continuation~\cite{Matcont}, as shown in Fig.~\ref{fig.1}c. 

The resulting ROM parameters, including effective mass $m_{\rm eff}^{(i)}$, $Q$-factor, linear stiffness $k^{(i)}=m_{\rm eff}^{(i)} (2 \pi f^{(i)})^2$ and nonlinear stiffness terms $a_{jk}$, $b_{jkl}$, are passed to the PSO optimizer (see Fig.~\ref{fig.1}d). The algorithm randomly generates many different initial designs by varying the geometric parameters, as shown in Fig.~\ref{fig.1}a. For each of these designs, known as a 'particle' in PSO, a ROM is generated by STEP and the corresponding objective functions are computed accordingly and passed to the optimizer. The optimizer then generates a next generation of particles based on the designs from the current generation, the objective functions, and the constraints, with the aim of improving their design parameters to optimize the objectives (see Supplementary Note 1). The optimization loop will iterate until it reaches the predefined maximum generation. 
If multiple objective functions are selected to be optimized, there is an additional step that selects the nondominated particles according to Pareto dominance~\cite{coello2004handling}. Because each particle is evaluated independently, PSO enables efficient parallel computing to evaluate all particles in one generation on a high-performance computing cluster. \\[1pt]

\begin{figure*}
\includegraphics[scale=0.85]{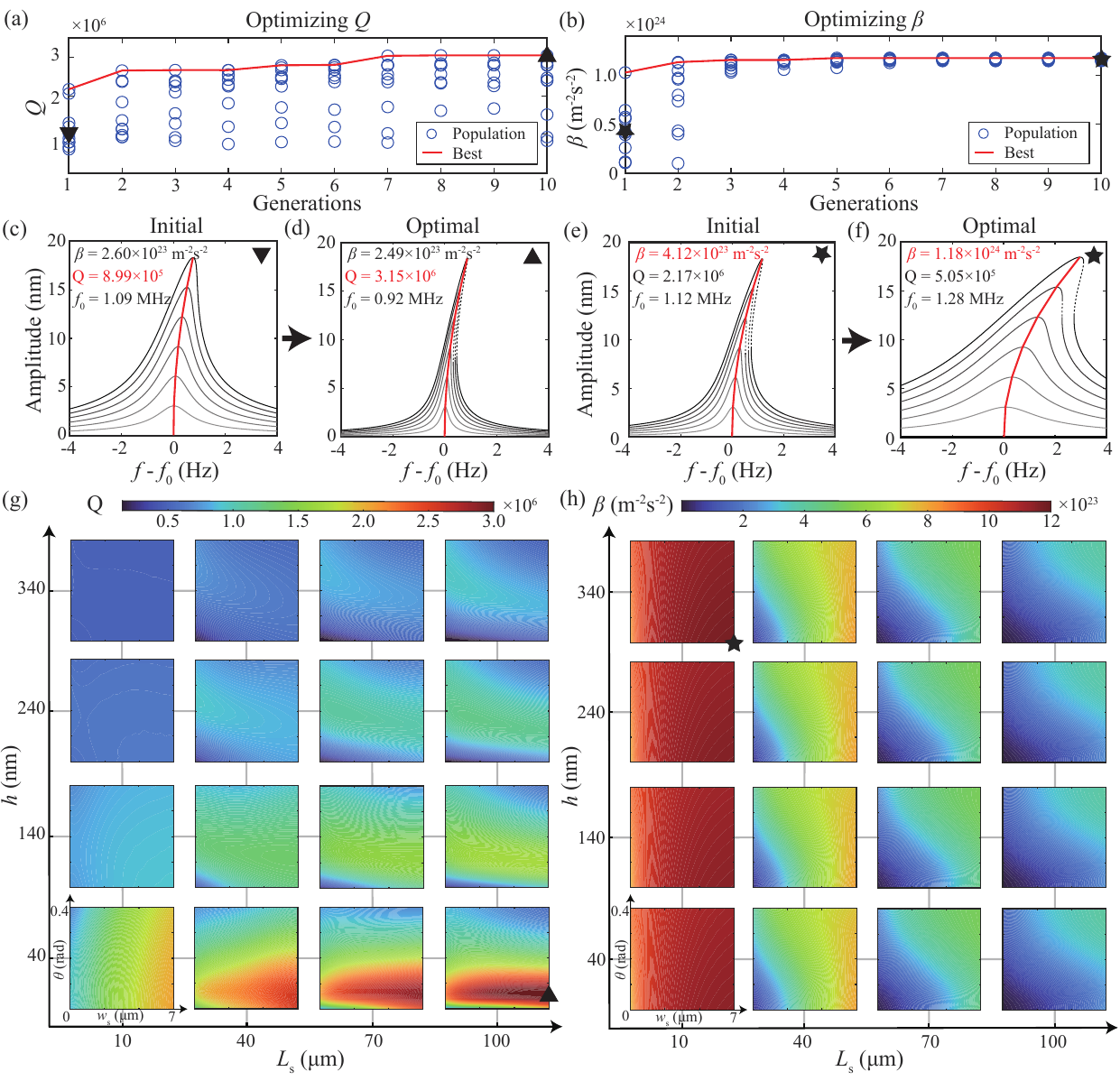}
\caption{\label{fig.2} \textbf{Optimal designs found by particle swarm optimization (PSO) and simulation of different dynamical properties.} Four geometric parameters are selected as design variables in Fig.~\ref{fig.1}. $w_{\rm s}$ and $\theta$ represent $x$ and $y$ axis, respectively, of each contour plot. (a,b) PSO's evolution shows the procedure of searching for maximum (a) $Q$ and (b) $\beta$, where the red lines mark the global best design of each generation. (c-f) Frequency response curves around the fundamental mode of (c,e) the designs with median performance in the initial generation and (d,f) the optimized designs, for $Q$ maximization (c,d) and for $\beta$ optimization (e,f), where the objectives and backbone curves are marked in red. The dotted lines are unstable solutions. The greyscale of response curves go from light to dark as the drive level increases. (g,h) Contour plots show the parametric study for (g) $Q$-factor and (h) mass-normalized Duffing constant $\beta$. The optimized designs found by PSO are marked as an upward-pointing triangle and a star, while the downward-pointing ones represent the designs with average objective values in the initial generation.} 
\end{figure*}

\section{OPTSTEP implementation and validation}
\subsection{Single objective optimization with OPTSTEP} 
We implement the presented OPTSTEP methodology to optimize the support geometry of the string resonator shown in Fig.~\ref{fig.1}a. The motion of the fundamental mode of the resonator can be described with the following nonlinear equation of motion:

\begin{equation}
\begin{aligned}
\ddot q + \frac{2 \pi f_0}{Q} \dot q + (2 \pi f_0)^2 q + \beta q^3 = F_{\rm exc} \sin{(2 \pi f t)},
\label{equation.2}
\end{aligned}
\end{equation}
\addtocounter{equation}{0}

\noindent where $q$ is the displacement at the string center, $f_0$ is the resonance frequency, $Q$ is the $Q$-factor, $\beta=b_{111}/m_{\rm eff}$ is the mass-normalized Duffing constant, and $F_{\rm exc} \sin{(2 \pi f t)}$ is the mass-normalized harmonic drive force. We present results of the OPTSTEP methodology for two optimization objectives, respectively: maximizing the $Q$-factor (shown in Fig.~\ref{fig.2}a,c,d) or maximizing the mass-normalized Duffing constant $\beta$ (shown in Fig.~\ref{fig.2}b,e,f) of the fundamental mode.
As design parameters, we use the support parameters ($L_{\rm s}$, $w_{\rm s}$, $\theta$ and $h$ in Fig.~\ref{fig.1}a). The PSO algorithm can freely initialize and vary these variables between preset constraints $\qty{10}{\micro\meter} <L_{\rm s}<\qty{100}{\micro\meter} $, $\qty{1}{\micro\meter} <w_{\rm s}<\qty{7}{\micro\meter} $, $\qty{0}{rad}< \theta<\qty{0.4}{rad}$, and $\qty{40}{\nano\meter} <h<\qty{340}{\nano\meter} $. 

We initialize the PSO algorithm with 10 randomly generated particles, as indicated by the blue circles at the first generation in Fig.~\ref{fig.2}a-b. The $Q$ and $\beta$ values of the best performing particle per generation are highlighted by the red line, which converges towards an optimum. 
Simulated response curves at different drive levels of the initial design (median performance of the initialized particles) and the optimized design are shown in Fig.~\ref{fig.2}c, d for $Q$ and Fig.~\ref{fig.2}e, f for $\beta$. It is obvious that the resonance peaks become narrower from Fig.~\ref{fig.2}c to Fig.~\ref{fig.2}d, indicative of an increase in $Q$-factor. From the backbone curves shown in Fig.~\ref{fig.2}e, f, we see that the resonance frequency of the optimized device shifts more at the same vibration amplitude, which suggests a larger, optimized value of $\beta$. 

\subsection{Numerical validation} 

In order to validate the PSO results, we compare them to a brute-force parametric study where we simulate a large number of designs that span the full design parameter space,  and plot the resulting values of $Q$ and $\beta$ in the contour plots in Fig.~\ref{fig.2}g, h.  Each of these subfigures consists of 16 small contour plots, each of which has a different combination of $L_{\rm s}$ and $h$, while along the axes the parameters $w_{\rm s}$ and $\theta$ are varied. The red-colored regions in the plots contain the optimal values of $Q$ and $\beta$, which are indicated by a triangle and a star. In Supplementary Table S1, we compare the optimized design parameters from the OPTSTEP method to the best devices from the parametric study. The close agreement between both approaches provides evidence that the OPTSTEP method is able to optimize both linear ($Q$) and nonlinear ($\beta$) parameters of the ROM. The results in Fig. 2a are obtained in 30 minutes using a high performance computing cluster, while the parametric study in Fig.~\ref{fig.2}g takes over 325 hours on the same cluster with the same amount of nodes. This illustrates the advantage in computation time that can be realized with OPTSTEP, although it is noted that these times strongly depends on the resolution of the parameter grid and other simulation parameters.


\begin{figure*}
\includegraphics[scale=0.82]{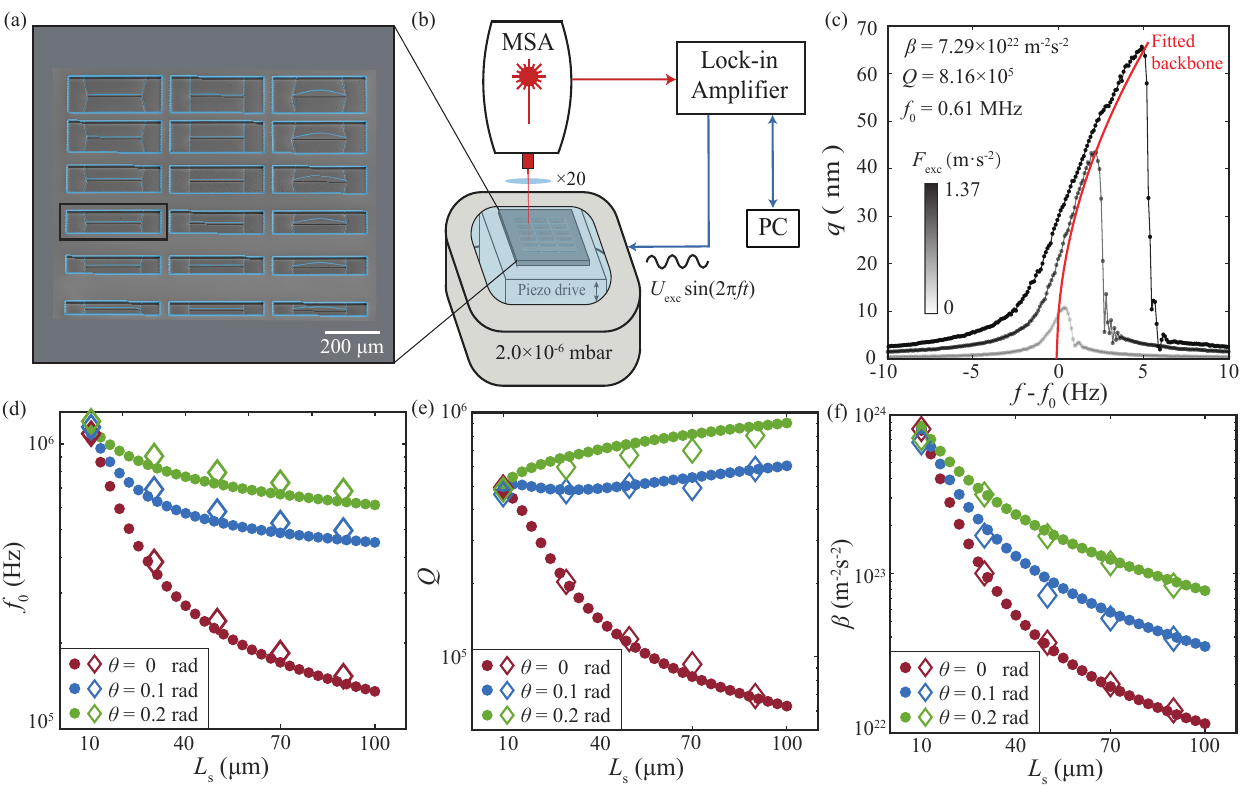}
\caption{\label{fig.3} \textbf{Experimental set-up and experimental validation of the simulations.} (a) Scanning Electron Microscope (SEM) image of an array of devices (colored in blue) with thickness $h = \qty[]{340}{\nano\metre}$ and different design variables. (b) Schematics of the measurement set-up, which includes a Micro System Analyzer (MSA) Laser Doppler Vibrometer (LDV) for motion detection and a piezo-actuator for driving the resonator. (c) Frequency response curves measured around the fundamental resonance frequency of the device with $L_{\rm s} = \qty[]{90}{\micro\metre}$, $w_{\rm s} = \qty[]{1}{\micro\metre}$, $\theta=\qty{0.20}{rad}$. The red curve is the fitted backbone. (d-f) Measured (diamonds) and FE-simulated (dots) resonance frequencies, $Q$-factor and Duffing constant $\beta$ for various values of the support length $L_{\rm s}$ and angle $\theta$, for devices with $w_{\rm s} = \qty[]{1}{\micro\metre}$ and $h = \qty[]{340}{\nano\metre}$. Error bars of measured results are smaller than the size of diamonds. }
\end{figure*}

\subsection{Experimental characterization} 

To compare the OPTSTEP method to experimental results, we also perform an experimental parametric study on 15 string resonators with varying support design parameters. For this we fabricated a set of devices with $\qty{10}{\micro\metre}<L_{\rm s}<\qty{90}{\micro\metre}$ and $\qty{0}{rad}<\theta<\qty{0.2}{rad}$, while keeping $h=\qty{340}{\nano\metre}$ and $w_{\rm s}=\qty{1.0}{\micro\metre}$ fixed. Fig.~\ref{fig.3}a shows a Scanning Electron Microscope (SEM) image of an array of nanomechanical resonators with varying support designs made of high-stress \ce{Si3N4} (see ``Methods'' for more details). To characterize the nonlinear dynamics of the devices, as shown in Fig.~\ref{fig.3}b, we fix the chip to a piezo actuator that drives the resonator by an out-of-plane harmonic base actuation in the out-of-plane direction. We use a Zurich Instruments HF2LI lock-in amplifier, connected to an MSA400 Polytec Laser Doppler Vibrometer, to measure the out-of-plane velocity at the center of the string resonator as a function of driving frequency (see Fig.~\ref{fig.3}c). We use a velocity decoder with a calibration factor of 200 mm/s/V. We perform all measurements in a vacuum chamber with a pressure below $\SI{2e-6}{\milli\bar}$ at room temperature.

\begin{figure*}
\includegraphics[scale=0.83]{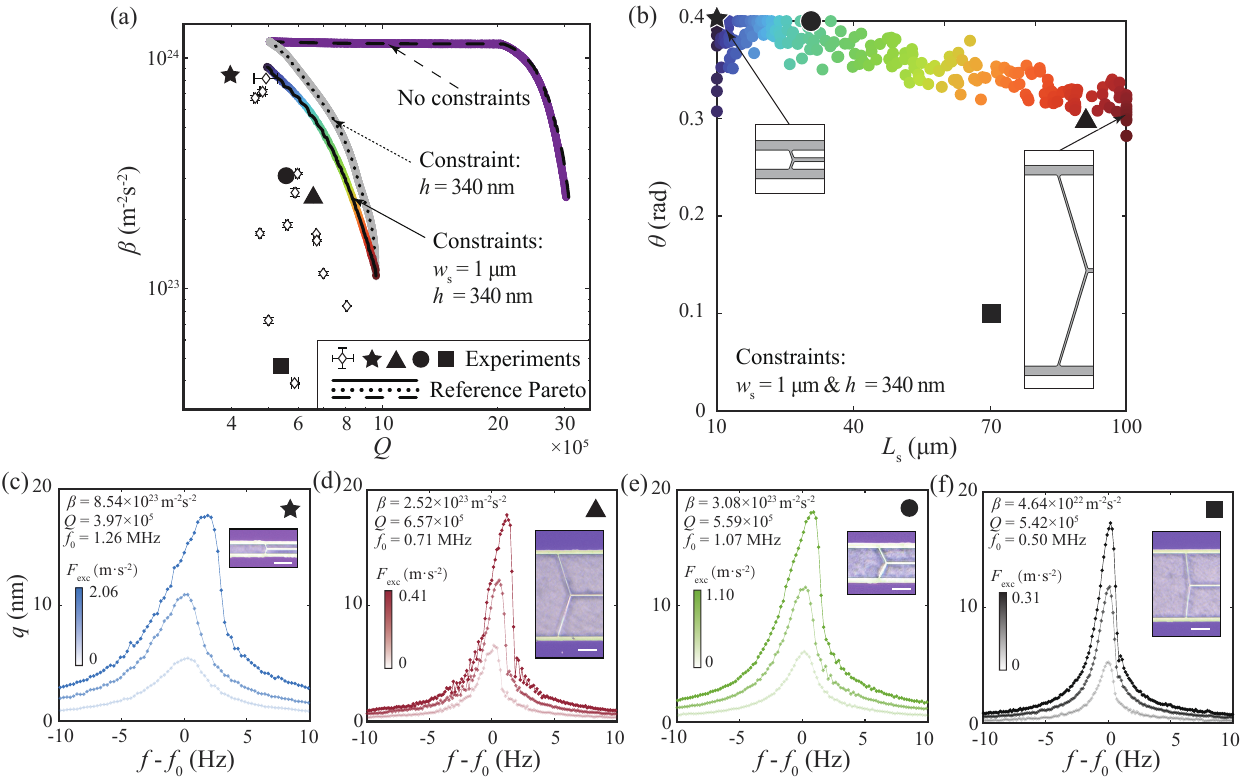}
\caption{\label{fig.4} \textbf{Trade-offs between $Q$-factor and the mass-normalized Duffing constant $\beta$ as obtained by combining OPTSTEP with multi-objective particle swarm optimization (MOPSO).} (a) Three Pareto fronts for different constraints (see main text) on design variables are shown in purple, grey and multi-colored dots. Measurements of devices that have the same design variables as the multi-colored Pareto front are shown by diamonds with error bars. The reference Pareto fronts (black solid, dotted and dashed lines) are generated by selecting the designs with maximum $Q$ and $\beta$ from the parametric study shown in Fig.~\ref{fig.2}g,h for the respective constraints (see Supplementary Note 2). (b) Each dot from the multi-colored Pareto front in (a) is plotted in the design space with the same color. The insets show the support design for a device with maximum $Q$ and a device with maximum $\beta$. (c-f) Measured frequency response curves for devices with maximum $\beta$ (c), maximum $Q$ (d), high $Q$ $\&$ $\beta$ (e), and low $Q$ $\&$ $\beta$ (f). Black symbols in the plots correspond with devices data points plotted in (a) and (b). The insets are images taken by Keyence digital microscope VHX-6000 and white scale bars are $\qty[]{20}{\micro\metre}$. 
}
\end{figure*}

Fig.~\ref{fig.3}c shows the frequency response at the center of the string at various drive levels for a device with $L_{\rm s} = \qty[]{90}{\micro\metre}$, $w_{\rm s} = \qty[]{1}{\micro\metre}$, $\theta=\qty{0.20}{rad}$ and $h = \qty[]{340}{\nano\metre}$. We estimate the linear resonator parameters of all devices by fitting the measured frequency response curves at various drive levels with the following harmonic oscillator function\cite{li2024strain}:  

\begin{equation}
\begin{aligned}
q_{\rm d} &= \frac{ q_{\rm max}/Q }{\sqrt{ \left[ 1- \left( f/f_0 \right)^2 \right]^2+f^2/(f_0 Q)^2}},
\label{equation.3}
\end{aligned}
\end{equation}

\noindent where $q_{\rm d}$ is the measured amplitude, $q_{\rm max}$ is the value of $q_{\rm d}$ when driving at the natural resonance frequency $f=f_0$, and $f$ is the drive frequency. To determine the nonlinear stiffness, we measure the resonator's frequency response at increasing drive levels, construct the backbone curve, and use the relation between the peak amplitude $q_\mathrm{max}$ and the peak frequency $f_{\rm max}$ to estimate the mass-normalized Duffing constant $\beta$ from\cite{nayfeh2008nonlinear,schmid2016fundamentals}:
\begin{equation}
\label{eq:backbone}
f_{\rm max}^2 = f_{\rm 0}^2 + \frac{3}{16 \pi^2} \beta q_{\rm max}^2.
\end{equation}
To compensate for small drifts in $f_0$ during the experiments, before fitting with Eq.~\eqref{eq:backbone}, we shift and align the frequency response curves to match their $f_0$ values\cite{li2024strain}.

In Fig.~\ref{fig.3}d-f, we compare the dynamical properties between FE-based ROMs (dots) and measurements on 15 string resonators (diamonds) as a function of $L_{\rm s}$ and $\theta$. It is evident that the fundamental resonance frequency $f_0$, $Q$-factor, and the mass-normalized Duffing constant $\beta$ of the fabricated devices, are all well predicted by FE-based ROMs. It can also be seen that for short support lengths $L_{\rm s}$ the device performance is similar, whereas increasing $L_{\rm s}$ allows tuning $f_0$, $Q$ and $\beta$ as we studied in more detail earlier\cite{li2023tuning, li2024strain}. 
In the next section we will compare these experimental results to multi-objective optimization as further validation of OPTSTEP.

\subsection{Multi-objective optimization with OPTSTEP}
For actual device design there are often multiple performance specifications that need to be met. It might sometimes be possible to condense these performance specifications into a single figure of merit, like the $f_0\times Q$ product for nanomechanical resonators. However, to make the best design decisions, it is preferred that the optimizer works with two (or more) objective functions like enhancing $f_0$ and $Q$, simultaneously. To enable this, we implement OPTSTEP with a multi-objective particle swarm optimization (MOPSO), which is an extension of single-objective PSO. After multi-objective optimization, the nondominated particles in the swarm are used to determine an approximation of the Pareto front, which is the set of designs for which improving one of the objectives will always lead to a deterioration of the other objective(s). By performing MOPSO, we aim at finding the Pareto front in the design space for multiple objectives, that represents the boundary on which all optimized designs reside for the chosen variables. As the red dots show in Fig.~\ref{fig.1}d illustrate, the Pareto front represents the boundary between feasible and unfeasible combinations of objectives and thus allows the designer to make the best trade-off among different objectives.

To demonstrate that multi-objective optimization can be combined with OPTSTEP, we use it to simultaneously maximize $Q$ and $\beta$. Devices with high quality factor and nonlinear stiffness can be of interest in cases where we are looking for designs that can drive a string into the nonlinear regime with a minimum driving force and power consumption. 

The resulting Pareto fronts are shown in Fig.~\ref{fig.4}a. Since we are also interested in the effect of the constraints on the optimum solutions, we include Pareto fronts with: no constraint (purple), a thickness constraint of $h = \qty[]{340}{\nano\metre}$ (grey), and with thickness and support width constraint (multi-coloured). These 3 Pareto fronts show that there is a clear trade-off between $Q$ and $\beta$, with higher $Q$-factor leading to lower nonlinearity $\beta$. The experimental devices share the same constraints ($w_{\rm s} = \qty[]{1}{\micro\metre}$ and $h = \qty[]{340}{\nano\metre}$) as the multi-colored Pareto and are plotted as the hollow diamonds with error bars in Fig.~\ref{fig.4}a (see Supplementary Table 2). We observe that all experimental points reside in the region on the left hand side of the Pareto front, confirming the area enclosed by the Pareto front indeed captures the feasible devices, and experimentally strengthening the confidence in the OPTSTEP approach for multi-objective designs. The color of the points links the points in the $Q-\beta$ graph in Fig.~\ref{fig.4}a to the corresponding design parameters in Fig.~\ref{fig.4}b. In Fig.~\ref{fig.4}b the schematic support geometries are shown as insets for both maximum $\beta$ (dark blue) and maximum $Q$ (dark red). We choose some of the fabricated devices close to the Pareto front to show typical measured frequency response curves and microscopic images in Fig.~\ref{fig.4}c-f, which correspond to the star, triangle, circle and square data markers in Fig.~\ref{fig.4}a and b. Together with the microscopic images, it is apparent that with minor alterations in the support region, the response of the string resonators can be largely tuned. To further explore the effect of other design parameters numerically, we release the constraint on $w_{\rm s}$, keeping only $h=\qty{340}{\nano\metre}$ constrained, and conduct MOPSO (see the grey Pareto front). We can see from the comparison between the grey and multicolored fronts that the performance gain from changing $w_{\rm s}$ is not very large. In contrast, if we further relax the constraint on $h = \qty[]{340}{\nano\metre}$, which shares the same design space in Fig.~\ref{fig.2}g-h, we obtain the purple Pareto front. The thinner $h$ pushes the Pareto front to have much higher $Q$. The long plateau at fixed $\beta$ is mainly attributed to the increase in $Q$ that results from the dependence of the intrinsic quality factor $Q_0$ on $h$ (see Methods). 
Besides validating the MOPSO approach by comparing with experimental data, we also use the data from the parametric study in Fig.~\ref{fig.2} to extract and generate reference Pareto fronts that are shown as black solid, dotted, and dashed lines in Fig.~\ref{fig.4}a (see Supplementary Note 2), with constraints that match those from the MOPSO optimization. \\[1pt]


\begin{figure*}
\includegraphics[scale=0.83]{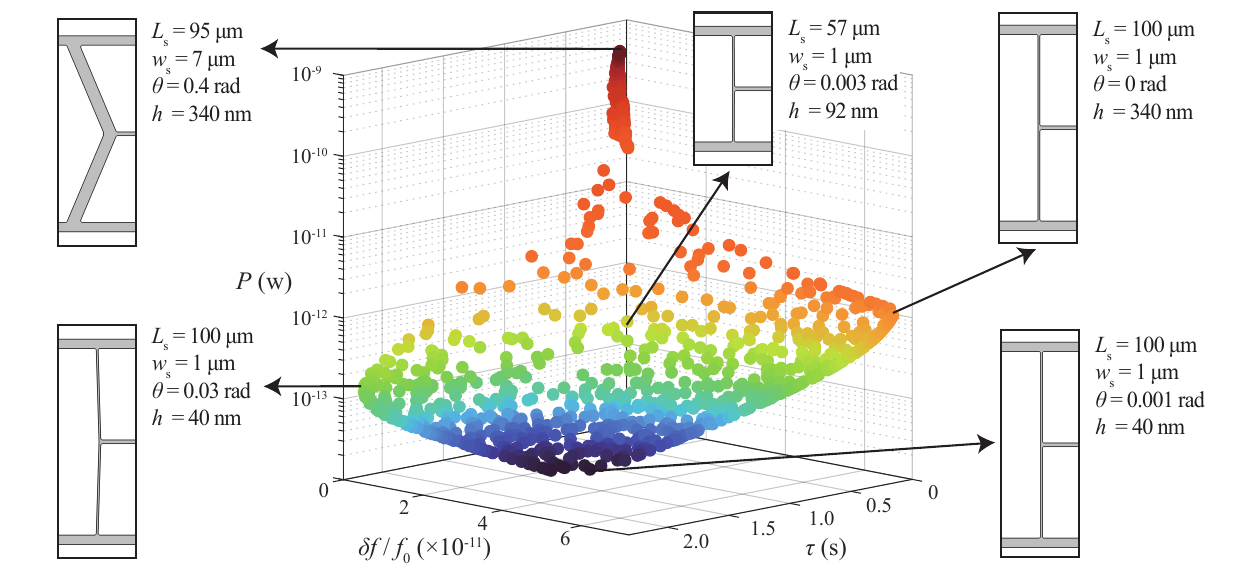}
\caption{\label{fig.5} \textbf{Trade-offs among the power consumption $P$, sensitivity $\delta f / f_0$ and response time $\tau$ of a string resonator with four design parameters.} The insets show the geometries and design parameters of supports of five representative designs on the Pareto frontier. The gradual change of color from dark blue to dark red marks the increasing in power consumption $P$ when operating the nanoresonator at the onset of nonlinearity $a_{\rm 1dB}$ to guarantee the maximum sensitivity.}
\end{figure*}

\section{Discussion}
The OPTSTEP methodology that is presented in this work enables the optimization of the nonlinear dynamic properties of resonant structures using standard FEM software, since it is based on the STEP and uses a derivative-free optimization method. We note that although derivative-free techniques like PSO are able to efficiently find near-optimal values of design parameters, optimality guarantees can typically not be given, and the techniques are therefore also called metaheuristic optimization techniques.
Here, in order to validate the OPTSTEP methodology numerically and experimentally, we have focused on $\beta$ and $Q$ maximization of the fundamental mode of a string resonator by geometric support design. After having established the methodology, it is now of interest to apply it to explore performance parameters that are more relevant to applications. For example, as shown in Fig.~\ref{fig.5}, our methodology can directly be extended to optimize the power consumption $P$, sensitivity $\delta f / f_0$ and response time $\tau$ of resonant sensors~\cite{demir2019fundamental,manzaneque2023resolution}, since these figure-of-merits can be directly expressed in terms of $m_{\rm eff}$, $f_0$, $Q$ and $\beta$ (see Supplementary Note 3). In Fig.~\ref{fig.5}, 1000 nondominated particles are found by OPTSTEP to form a 3D surface that approaches the Pareto frontier with the objective of minimizing $P$, $\delta f / f_0$ and $\tau$ simultaneously. The particles have the same design constraints as in the example in Fig.2 and the purple Pareto front in Fig.~\ref{fig.4}a, which are $\qty{10}{\micro\metre}<L_{\rm s}<\qty{90}{\micro\metre}$ and $\qty{0}{rad}<\theta<\qty{0.2}{rad}$ and membrane thickness $\qty{40}{\nano\metre} <h<\qty{340}{\nano\metre}$. 
The competing design trade-offs between these three objective functions are obtained from OPTSTEP, and are visualized in Fig.~\ref{fig.5} by showing five typical designs near the Pareto frontier. As demonstrated by the designs at the upper right corner of the Pareto frontier, we can conclude that the devices with shorter response time are more likely to have thicker supports, which lead to a higher resonance frequency $f_0$ combined with a low $Q$, thus resulting in a smaller $Q / f_0$ ratio. At the same time, these thicker supports also contribute to a larger onset of nonlinearity $a_{\rm 1dB}$~\cite{li2024strain}, so the resonators are able to work at much larger amplitudes in the linear regime, which provides a better sensitivity $\delta f / f_0$. However, the larger $a_{\rm 1dB}$ and $m_{\rm eff}$ will require more energy to sustain the oscillation at resonance that causes higher power consumption $P$. In contrast, the devices with higher sensitivity $\delta f / f_0$, which are shown at the lower left corner in Fig.~\ref{fig.5}, are equipped with more slender supports. With only a slight increase of support angle $\theta$ from 0, the low torsional stiffness of supports is maintained while the stress in the central string can be significantly increased~\cite{li2023tuning}, leading to a higher $Q$, which can be confirmed by Fig.~\ref{fig.2}g. Consequently, when aiming at designing a resonant sensor with relatively low power consumption $P$, high sensitivity $\delta f / f_0$ and short response time $\tau$ with compliant supports, a pair of slender and slightly angled supports, together with a medium thickness of \ce{Si3N4} layer is generally favored.

In other cases, like approaching the quantum regime with a nonlinear nanomechanical resonator~\cite{samanta2023nonlinear}, it is beneficial to maximize $Q$ and $\beta$ simultaneously. The OPTSTEP methodology can also be used for more complex design problems that involve multiple modes~\cite{Dou2015,foster2016tuning,miller2021amplitude,li2024strain}, for avoiding or taking advantage of mode coupling, for instance by optimizing nonlinear coupling coefficients ($a_{jk}$ and $b_{jkl}$ in Fig.~\ref{fig.1}b) and resonance frequency ratios. Since OPTSTEP generates the ROM parameters at each generation, it is particularly suited for dealing with cases where the device specifications can be expressed in terms of these parameters. Interesting challenges include increasing frequency stability by coherent energy transfer~\cite{Daniel2012frequencystablization,Chen2017coherenttime}, signal amplification~\cite{badzey2005coherent} and stochastic sensing~\cite{venstra2013stochastic, belardinelli2023hidden}. Moreover, intriguing paths for further research involve inclusion of nonlinear damping or extension to full topology optimization~\cite{hoj2021ultra}. Also the use of alternative optimization strategies, like binary particle swarm optimization (BPSO)~\cite{lake2013particle}, that could generate radically new geometries, is an interesting direction.

\section{Conclusions}
To sum up, we presented a methodology (OPTSTEP) for optimizing the nonlinear dynamics of mechanical structures by combining an FE-based ROM method with a derivative-free optimization technique (PSO). We demonstrated and validated the methodology by optimizing the support design of high-stress \ce{Si3N4} nanomechanical resonators. The method was verified numerically by comparing its results to a brute-force parametric study, for both single- and multi-objective optimization. Experimental data on the $Q$-factor and Duffing nonlinearity were in correspondence with the OPTSTEP results. The capability of the method was also demonstrated by multi-objective optimization of the support for the nanomechanical resonator, targeting improvements in power consumption, sensitivity and response time in resonant sensing. 
We thus concluded that the method can be applied to a wide range of complex design challenges including nonlinear dynamics, and is expected to be compatible to most FE codes and derivative-free optimization routines. It holds the potential to facilitate and revolutionize the way (nano)dynamical systems are designed, thus pushing the ultimate performance limits of sensors, mechanisms and actuators for scientific, industrial, and consumer applications. \\[1pt]

\section{Methods}
\noindent{\textbf{Sample fabrication.} We produce our nanomechanical resonators using electron beam lithography and reactive ion etching techniques on high-stress \ce{Si3N4} layers, chosen for their reliability and precision in achieving design specifications~\cite{Spiderweb}. These layers are deposited via low pressure chemical vapor deposition (LPCVD) onto a silicon substrate. Following this, the devices undergo suspension through a fluorine-based deep reactive ion underetching process. The mechanical properties of the high-stress \ce{Si3N4} are characterized in our previous works~\cite{li2024strain}, with an initial isotropic stress $\sigma_0 = \SI{1.06}{\giga\pascal}$, Young's modulus $E = \SI{271}{\giga\pascal}$, Poisson's ratio $\nu=0.23$, mass density $\rho = \SI{3100}{\kilogram/\metre^3}$. The intrinsic quality factor is a function of thickness $h$~\cite{villanueva2014evidence}, which is $Q_0^{-1} = 28000^{-1} + \left(6 \times 10^{10} h \right)^{-1}$. \\[1pt]

\section{Data availability}
The data that support the findings of this study are available from the corresponding authors upon reasonable request. \\[1pt]


\section{Acknowledgements}
Funded/Co-funded by the European Union (ERC Consolidator, NCANTO, 101125458). Views and opinions expressed are however those of the author(s) only and do not necessarily reflect those of the European Union or the European Research Council. Neither the European Union nor the granting authority can be held responsible for them. Z.L. acknowledges financial support from China Scholarship Council, the assistance on the FE reduced-order modeling from Vincent Bos, and the instruction about using the high
performance computing cluster from Binbin Zhang. This work is also part of the project, Probing the physics of exotic superconductors with microchip Casimir experiments (740.018.020) of the research program NWO Start-up which is partly financed by the Dutch Research Council (NWO). M.X. and R.A.N. acknowledge valuable support from the Kavli Nanolab Delft. \\[1pt]

\section{Author contributions}
Z.L., F.A., P.G.S. and A.M.A. conceived the experiments and methods; M.X. and R.A.N. fabricated the \ce{Si3N4} samples; Z.L. conducted the measurements and analysed the experimental data; Z.L. and F.A. built the theoretical model; Z.L. performed the reduced-order modelling of the finite element model; Z.L. and A.S. set up the optimization on high performance cluster; F.A. and P.G.S. supervised the project; and the manuscript was written by Z.L. and P.G.S. with inputs from all authors. \\[1pt]

\section{Competing interests}
The authors declare no competing interests}.

\bibliography{apssamp}

\clearpage



\section*{Supplementary material}

# Finite Element-based Nonlinear Dynamic Optimization of Nanomechanical Resonators

Zichao Li[*, 1], Farbod Alijani[1], Ali Sarafraz[1], Minxing Xu[1, 2], Richard A. Norte[1, 2], Alejandro M. Aragón[1], and Peter G. Steeneken[*, 1, 2]

[1]Department of Precision and Microsystems Engineering, Delft University of Technology, Mekelweg 2, 2628 CD Delft, The Netherlands
[2]Kavli Institute of Nanoscience, Delft University of Technology, Lorentzweg 1, 2628 CJ Delft, The Netherlands

July 10, 2024

## Supplementary Note 1: Particle swarm optimization

Here, we elaborate the details in our optimizer as show in Supplementary Fig. 1, which is based on the algorithms initiated by [1, 2]. For single objective function, the velocity of the $i$th particle of the $(j+1)$th generation is decided by three parts respectively [1]:

$$v_i^{(j+1)} = C_0 v_i^{(j)} + C_1 \left(PBest_i^{(j)} - x_i^{(j)}\right) + C_2 \left(GBest^{(j)} - x_i^{(j)}\right), \tag{S1}$$

where $v_i^{(j)}$ and $x_i^{(j)}$ are the velocity and position of the $i$th particle at the $j$th generation, $PBest_i^{(j)}$ is the best design that the $i$th particle has had, $GBest^{(j)}$ is the best design among particles in the $j$th generation. $C_0$ is the inertia weight, $C_1$ and $C_2$ are acceleration coefficients. $C_0$, $C_1$ and $C_2$ are all set as 2. Accordingly, we can update the design of the $i$th particle at the $(j+1)$th generation as:

$$x_i^{(j+1)} = x_i^{(j)} + v_i^{(j+1)}. \tag{S2}$$

In order to deal with multiple objectives, the concept of Pareto dominance is introduced [2]. A particle is Pareto optimal (nondominated) if there exists no other particles in the same generation that would improve some objectives without causing a simultaneous derogation in at least one other objective. According to the Pareto dominance, we selected nondominated particles from each generation and stored them in an external repository, which is updated for each iteration. The velocity of the $i$th particle of the $(j+1)$th generation is derived based on:

$$v_i^{(j+1)} = C_0 v_i^{(j)} + C_1 \left(PBest_i^{(j)} - x_i^{(j)}\right) + C_2 \left(Rep^{(j)} - x_i^{(j)}\right), \tag{S3}$$

where $Rep^{(j)}$ is one nondominated particle chosen from the repository. The design space is divided into hypercubes and the nondominated particles fall into different hypercubes randomly. The hypercubes with more nondominated particles are assigned a lower weight that are less likely to be selected according to roulette-wheel selection. This promotes a more equidistantly distributed nondominated particles along the Pareto front.

Besides, to reduce the drawbacks of PSO that it may converge to a local minimum due to a high convergence speed, a mutation operator is introduced after the design update of all particles. The mutation rate is designed to drop as the generation increases, which guarantees a high exploration of the whole design space at the beginning of the optimization.

In Supplementary Table.1, we list the values and corresponding geometric design parameters of both $Q$-factor and the mass-normalized Duffing constant $\beta$ obtained by PSO and by searching the parametric study results in Fig. 2 of the main text.

Supplementary Table 1: Values and corresponding geometric design parameters for $Q$ and $\beta$ in Fig. 2

| Dynamical properties | Values | $L_s$ (µm) | $w_s$ (µm) | $\theta$ (rad) | $h$ (nm) |
|---|---|---|---|---|---|
| Optimized $Q$ from PSO | $3.05 \times 10^6$ | 83.51 | 6.87 | 0.06 | 40.00 |
| Maximum $Q$ from parametric study | $3.15 \times 10^6$ | 100.00 | 7.00 | 0.06 | 40.00 |
| Optimized $\beta$ from PSO | $1.18 \times 10^{24} (\mathrm{m^{-2} s^{-2}})$ | 10.00 | 7.00 | 0.03 | 340.00 |
| Maximum $\beta$ from parametric study | $1.18 \times 10^{24} (\mathrm{m^{-2} s^{-2}})$ | 10.00 | 7.00 | 0 | 340.00 |

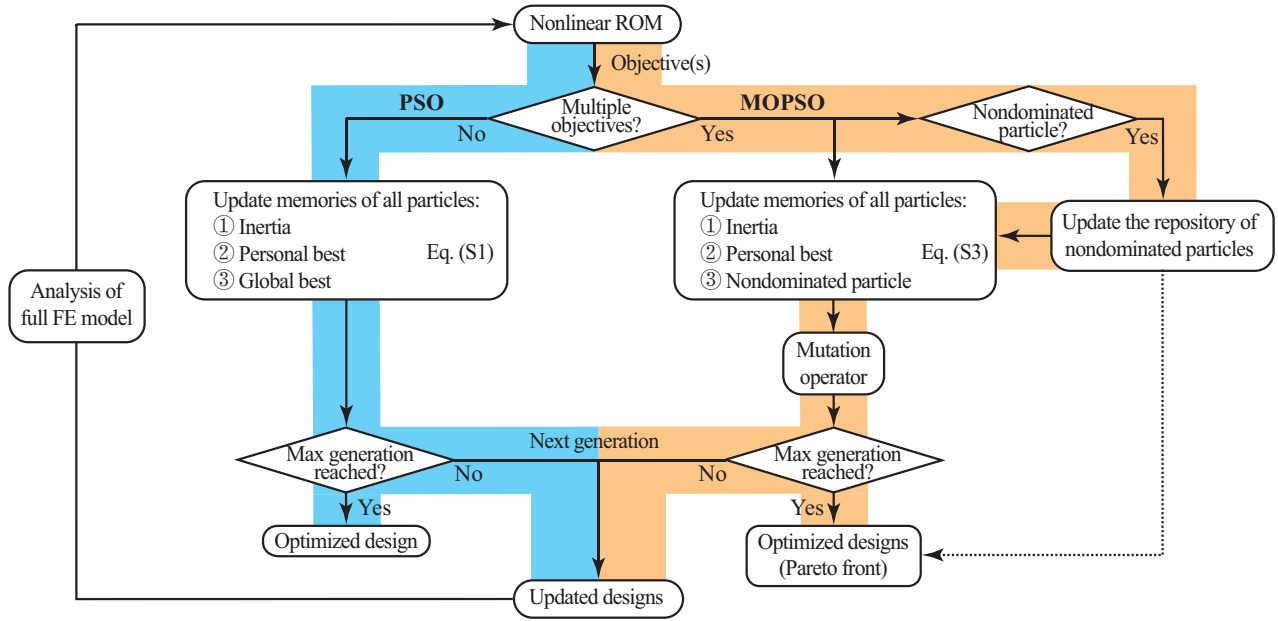

Supplementary Figure 1: **Details of the optimizer in Fig. 1.** According to the amount of objective functions, two routes based on particle swarm optimization (PSO) are available, respectively. A repository for nondominated particles is built additionally for multi-objective particle swarm optimization (MOPSO).

## Supplementary Note 2: Validation of MOPSO

We generate reference Pareto fronts from parametric studies according to the Pareto criterion mentioned in Supplementary Note 2, as shown in Supplementary Fig. 2. We evenly discretize the design space $L_s = 10 \sim 100\,\mu m$, $\theta = 0 \sim 0.4\,\mathrm{rad}$, $w_s = 1 \sim 7\,\mu m$ and $h = 40 \sim 340\,\mathrm{nm}$ and build the 1-dof nonlinear reduced-order model according to the finite element analysis of the full model. The black lines are used in Fig. 4(a) of the main text as references for the MOPSO generated Pareto fronts.

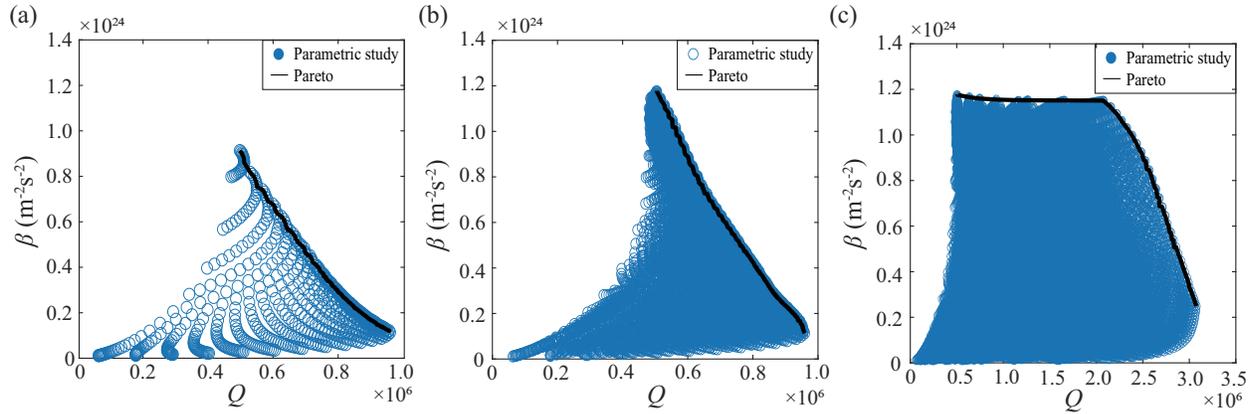

Supplementary Figure 2: **Reference Pareto fronts of different design spaces.** (a) Pareto front of varying $L_s$ and $\theta$ while fixing $w_s = 1\,\mu m$ and $h = 340\,\mathrm{nm}$. (b) Pareto front of varying $L_s$, $\theta$ and $w_s$ while fixing $h = 340\,\mathrm{nm}$. (c) Pareto front of varying $L_s$, $\theta$, $w_s$ and $h$. The blue circles show the FE-based ROMs of devices corresponding to different design spaces.

Besides, we show the measured $Q$-factor and the mass-normalized Duffing constant $\beta$ of our fabricated devices with constraints $w_s = 1\,\mu m$ and $h = 340\,\mathrm{nm}$. We observe that they all reside in the lower left side compared to the colored Pareto front, which proves the reliability of Pareto fronts found by MOPSO. In Supplementary Table.2, we list the geometric parameters of the measured devices, which are shown as diamonds with error bars in Fig. 4(a), and their corresponding $Q$ and $\beta$. Some of the devices listed in Supplementary Table.2 is out of the range of Fig. 4(a) because of low $Q$ or $\beta$.

Supplementary Table 2: Values and corresponding geometric design parameters for $Q$ and $\beta$ in Fig. 4(a)

| $\theta$ (rad) | $L_\mathrm{s}$ (μm) | $Q$ | $\beta$ (m$^{-2}$s$^{-2}$) | Mark |
|---|---|---|---|---|
| 0.4 | 90 | 671081 | $1.62 \times 10^{23}$ | |
| 0.4 | 70 | 561570 | $1.89 \times 10^{23}$ | |
| 0.4 | 50 | 588671 | $2.61 \times 10^{23}$ | |
| 0.4 | 30 | 558798 | $3.08 \times 10^{23}$ | ● |
| 0.4 | 10 | 397014 | $8.54 \times 10^{23}$ | ★ |
| 0.3 | 90 | 657047 | $2.52 \times 10^{23}$ | ▲ |
| 0.2 | 90 | 804593 | $8.44 \times 10^{22}$ | |
| 0.2 | 70 | 698133 | $1.17 \times 10^{23}$ | |
| 0.2 | 50 | 668694 | $1.73 \times 10^{23}$ | |
| 0.2 | 30 | 597314 | $3.15 \times 10^{23}$ | |
| 0.2 | 10 | 483734 | $7.15 \times 10^{23}$ | |
| 0.1 | 90 | 587378 | $3.90 \times 10^{22}$ | |
| 0.1 | 70 | 541695 | $4.64 \times 10^{22}$ | ■ |
| 0.1 | 50 | 501157 | $7.30 \times 10^{22}$ | |
| 0.1 | 30 | 475866 | $1.74 \times 10^{23}$ | |
| 0.1 | 10 | 462534 | $6.71 \times 10^{23}$ | |
| 0 | 90 | 69112 | $1.38 \times 10^{22}$ | |
| 0 | 70 | 93174 | $2.02 \times 10^{22}$ | |
| 0 | 50 | 118166 | $3.68 \times 10^{22}$ | |
| 0 | 30 | 202588 | $1.01 \times 10^{23}$ | |
| 0 | 10 | 493902 | $8.15 \times 10^{23}$ | |

## Supplementary Note 3: Figure-of-merits in a resonant sensor

Firstly, the response time $\tau$ demonstrates the responsivity of a resonant sensor subjected to external stimuli, which is related to the resonance frequency and the corresponding $Q$-factor [3]:

$$\tau = \frac{Q}{\pi f_0} \tag{S4}$$

Secondly, for a resonant sensor operate in linear regime, the sensitivity is determined by its frequency stability that is influenced by noise. The highest signal-to-noise ratio can be obtained by operating it at the onset of nonlinearity $a_\mathrm{1dB}$, where it has the largest amplitude [3, 4, 5]. Assuming the thermomechanical noise dominates the noise floor, we use the expression of frequency stability in the closed loop measurement to evaluate the sensitivity of our devices:

$$\frac{\delta f}{f_0} \propto \frac{1}{2Q} 10^{\mathrm{DR}/20}, \tag{S5}$$

where DR can be expressed as the function of the ratio between the onset of nonlinearity $a_\mathrm{1dB}$ and the thermomechanical noise $a_\mathrm{th}$ [6]:

$$\begin{aligned}
\mathrm{DR} &= 20\lg \frac{a_\mathrm{1dB}}{a_\mathrm{th}} \\
&= 20\lg \frac{0.9244\sqrt{\frac{\alpha}{Q\beta}}}{\sqrt{\frac{4k_\mathrm{B} T Q \Delta f}{m_\mathrm{eff} \alpha^{\frac{3}{2}}}}} \\
&= 20\lg \left[ 0.4622 \left(k_\mathrm{B} T \Delta f\right)^{-\frac{1}{2}} m_\mathrm{eff}^{-\frac{1}{2}} Q^{-1} \alpha^{\frac{5}{4}} \beta^{-\frac{1}{2}} \right].
\end{aligned} \tag{S6}$$

$k_\mathrm{B}$ is the Boltzmann's constant. We assume the resonator operates in room temperature $T = 298$K with the measurement bandwidth $\Delta f = 1$Hz. After substituting Eq.S6 into Eq.S5, we can have the relationship between $\delta f/f_0$ and other parameters from ROM, which is irrelevant to $Q$:

$$\frac{\delta f}{f_0} \propto m_\mathrm{eff}^{\frac{1}{2}} \alpha^{-\frac{5}{4}} \beta^{\frac{1}{2}}. \tag{S7}$$

Thirdly, we define the power consumption $P$ as the energy that is needed to be fed to the devices per second, i.e., the energy dissipated per second:

$$P = \frac{\Delta W}{T} = \frac{f_0}{2\pi} \Delta W. \tag{S8}$$

Considering the definition of damping ratio:

$$\zeta = \frac{1}{2Q} = \frac{\Delta W}{4\pi W}, \tag{S9}$$

and the total energy stored in the resonator supposed that it is operated at the onset of nonlinearity $a_{1\text{dB}}$ for the best sensitivity:

$$W = \frac{1}{2} m_{\text{eff}} \alpha a_{1\text{dB}}^2, \tag{S10}$$

we can derive the power consumption $P$ as:

$$P = \frac{f_0}{Q} \cdot \frac{1}{2} m_{\text{eff}} \alpha a_{1\text{dB}}^2. \tag{S11}$$

It is worth noticing that when the resonator is operated at the onset of nonlinearity, the potential energy stored in the geometric nonlinearity is several magnitudes smaller than $W$, that can be neglected.

## Supplementary References



\end{document}